\documentclass{internoise2026}

\fancypagestyle{plain}{%
  \fancyhf{}

}
\fancypagestyle{empty}{%
  \fancyhf{}

}
\makeatletter
\def\@oddhead{}
\def\@evenhead{}
\def\@oddfoot{}
\def\@evenfoot{}
\makeatother

\begin{document}
\thispagestyle{empty}

\begin{center}
\vspace*{-0.5cm}

{\Large \bfseries
Transformer-based End-to-End Control Filter Generation\\
for Active Noise Control\footnote{This is the author's preprint version of a manuscript submitted to INTER-NOISE 2026.}
\par}

\vspace{0.45cm}

{\normalsize
Ziyi Yang, Zhengding Luo,Yisong Zou, Boxiang Wang, Qirui Huang, and Woon-Seng Gan
\par}

\vspace{0.6cm}
\end{center}

\vspace{0.6cm}

\begin{abstract}
To address the limitations of existing Generative Fixed-Filter Active Noise Control (GFANC) methods, which rely on filter decomposition and recombination and require supervised learning with labeled data, this paper proposes a Transformer-based End-to-End Control-Filter Generation (E2E-CFG) framework. Unlike previous approaches that predict combination weights of sub control filters, the proposed method directly generates control filters in an unsupervised manner by integrating the co-processor and real-time controller into a fully differentiable ANC system, where the accumulated error signal is used as the training objective. By abandoning the decomposition--reconstruction process, the proposed design simplifies the control pipeline and avoids error accumulation, while the Transformer architecture effectively captures global and dynamic noise characteristics through its attention mechanism. Numerical simulations on real-recorded noises demonstrate that the proposed method achieves improved noise reduction performance and adaptability to different types of noises compared with the original GFANC framework.
\end{abstract}

\section{Introduction}

Active noise control (ANC) reduces unwanted sound through destructive interference and has been studied for decades in applications such as ducts, vehicles, headsets, window, and enclosed spaces \cite{burgess1981active,nelson1992active,yang2023slit}. Classical feedforward and feedback ANC systems, especially those based on the filtered-x least mean square (FxLMS) algorithm and its variants, remain fundamental because of their clear physical interpretation and practical effectiveness \cite{kuo1999active,ji2025mixed,kajikawa2012recent,li2024experimental}. However, adaptive ANC is still sensitive to causality constraints, secondary-path modeling, and convergence behavior, which limits its performance in rapidly changing acoustic conditions, particularly when low-latency implementation is required \cite{elliott2001signal}.

As an alternative to continuously adapting controllers, fixed-filter ANC has been widely considered in practical systems because it provides immediate response without the slow convergence associated with online adaptation \cite{iotov2022computationally,iotov2023nonstationary,shen2022adaptivegain}. Based on this idea, selective fixed-filter active noise control (SFANC) was proposed to select a suitable pre-trained filter according to the incoming noise condition, and later learning-based SFANC methods used convolutional neural networks (CNNs) to automate filter selection and improve practicality \cite{shi2020sfanc,shi2022cnnsfanc,luo2022hybridsfanc}. While these methods improve the flexibility of fixed-filter ANC, selecting one candidate from a limited filter set may still be insufficient when the incoming noise differs substantially from the design conditions.

To address this limitation, generative fixed-filter active noise control (GFANC) was developed to generate a more suitable control filter by combining sub-control filters rather than selecting only one candidate \cite{luo2024bayes}. Existing studies showed that this strategy improves the adaptability of fixed-filter ANC, and temporal smoothing mechanisms such as Bayesian or Kalman filtering can further enhance robustness under dynamic noise conditions \cite{luo2024bayes,luo2024kalman}. Nevertheless, current GFANC frameworks still generate the control filter indirectly through a decomposition-and-recombination process, which increases pipeline complexity and makes performance dependent on the intermediate filter representation. In addition, the co-processor is typically trained in a supervised manner, requiring labelled targets and extra offline data preparation. Recently, Luo \textit{et al.} proposed an unsupervised GFANC framework, where the co-processor and real-time controller are integrated into a differentiable ANC system and the accumulated squared error is used directly as the training objective \cite{luo2024unsupervised}. This result suggests that GFANC can be trained without labelled data while remaining directly aligned with the physical objective of noise cancellation.

In parallel, neural-network-based ANC has been increasingly explored beyond conventional adaptive filtering \cite{zhang2021deepanc,zhang2023arn,bai2025wavenet,wang2025transferable}. Under the end-to-end unsupervised GFANC formulation, the remaining question is whether the sequence modeling capability of the co-processor can be further improved. Existing GFANC methods mainly rely on one-dimensional CNNs, which are efficient but primarily operate through local receptive fields \cite{luo2024unsupervised}. For control-filter generation, however, the appropriate filter may depend on more time-varying characteristics of the incoming noise. Transformer architectures provide a different mechanism through self-attention and have shown strong performance in sequential speech and audio tasks \cite{vaswani2017attention,gulati2020conformer,subakan2021sepformer}. Motivated by these results, this paper investigates a Transformer-based End-to-End Control-Filter Generation (E2E-CFG) framework. Building on the unsupervised GFANC formulation, we replace the CNN-based co-processor with a Transformer-based architecture and directly generate the control filter within a fully differentiable ANC system. Numerical simulations on real-recorded noises show that the proposed method achieves improved noise reduction performance and better adaptability across different noise types than the original GFANC framework. The main contributions of this work are:
\begin{itemize}[label=\textbullet]
\item \textbf{End-to-end control-filter generation}: we develop a differentiable ANC system in which the control-filter coefficients are directly generated for each input noise frame, without relying on sub-filter decomposition and recombination, thereby reducing the gap between the generated filters and the optimal control filters.
\item \textbf{Transformer-based unsupervised learning}: we introduce a Transformer-based co-processor and train it in an unsupervised manner by directly minimizing the accumulated residual error, without requiring labelled target filters.
\item \textbf{Generalization to unseen noises}: under the same end-to-end training paradigm and with training conducted only on synthetic noises, we compare the proposed method with the previous GFANC baseline \cite{luo2024unsupervised} and show that the proposed method yields more consistent improvement on unseen real-noise conditions.
\end{itemize}

\section{Proposed E2E-CFG Framework}

This paper proposes a Transformer-based End-to-End Control-Filter Generation (E2E-CFG) framework. The proposed method differs from previous GFANC approaches \cite{luo2025deep,luo2024unsupervised} in two aspects. First, the CNN-based co-processor is replaced by a Transformer-based architecture for frame-wise control-filter generation. Second, instead of generating the final control filter through sub-filter decomposition and recombination, the proposed method directly predicts the control-filter coefficients from buffered input frames. The framework follows a two-rate structure: the physical ANC path operates at the sampling rate, while the neural co-processor updates the control filter at the frame rate.

\subsection{Overall framework}

\begin{figure}[htbp!]
  \centering
  \includegraphics[width=0.8\linewidth]{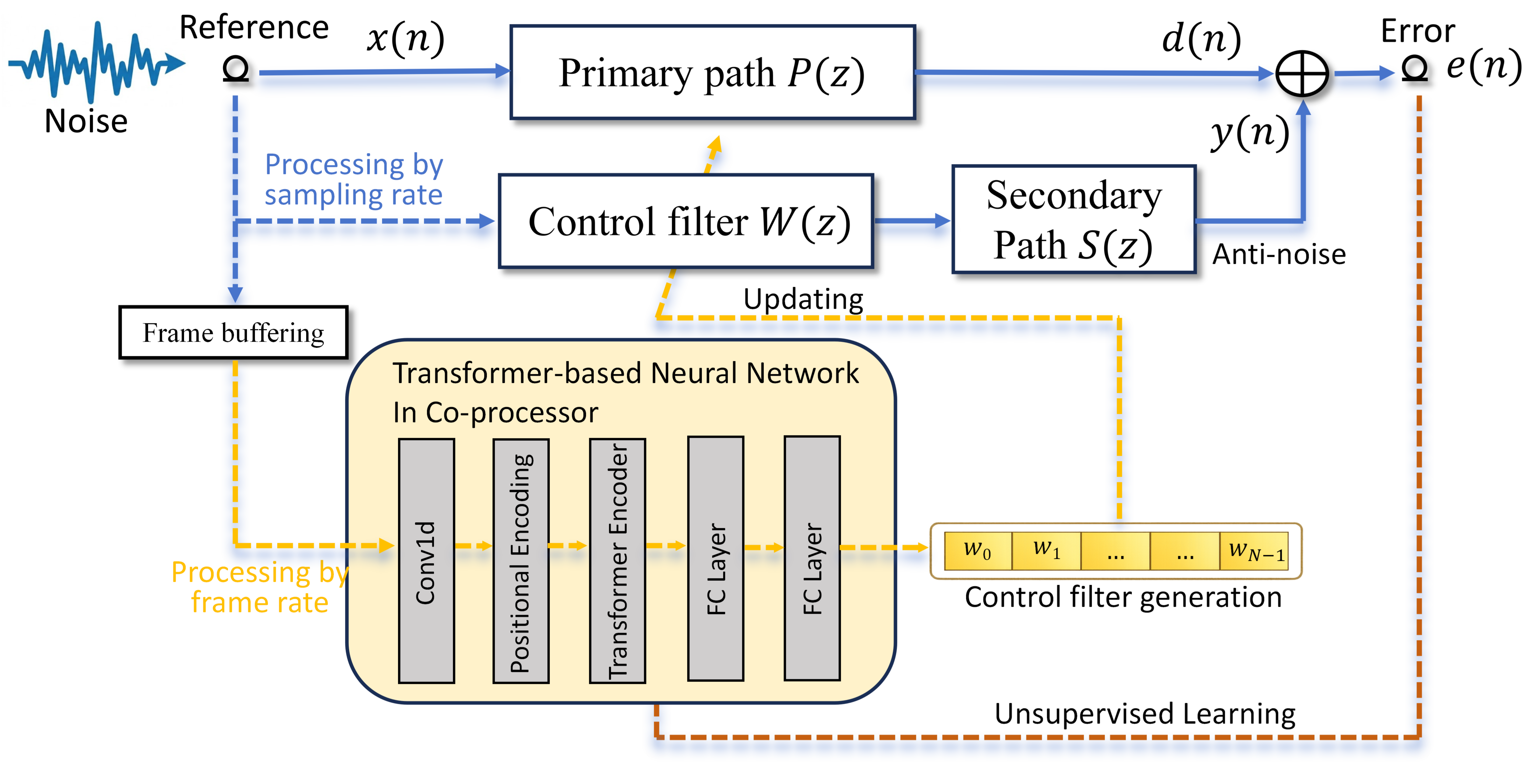}
  \caption{Overview of the proposed Transformer-based End-to-End Control-Filter Generation framework.}
  \label{fig:framework}
\end{figure}

Figure~\ref{fig:framework} illustrates the proposed framework. Let $x(n)$ denote the reference signal, where $n$ is the discrete-time sample index. Through the primary path $P(z)$, the disturbance signal at the error position is denoted by $d(n)$. In parallel, the control filter $W(z)$ drives the secondary path $S(z)$ to produce the anti-noise signal $y(n)$. The residual error is
\begin{equation}
e(n)=d(n)-y(n),
\label{eq:error_signal}
\end{equation}
where $e(n)$ denotes the remaining noise after cancellation.

Instead of updating the controller sample by sample, the proposed method buffers the reference signal into frames and feeds each frame to a Transformer-based co-processor. For every input frame, the co-processor generates one control-filter coefficient vector, which is then assigned to the real-time controller. In this way, the physical ANC path remains sample-wise, while the control filter is updated frame-wise.

\subsection{Transformer-based co-processor}

Let $\mathbf{x}_{f}\in\mathbb{R}^{L}$ denote one buffered input frame of length $L$. The proposed co-processor consists of a Conv1d layer, positional encoding, a Transformer encoder, and two fully connected (FC) layers. The Conv1d layer extracts local temporal patterns from the waveform, while the Transformer encoder captures longer-range temporal dependencies within the frame. The network output is the control-filter coefficient vector
\begin{equation}
\mathbf{w}=\mathcal{F}_{\theta}(\mathbf{x}_{f}),
\label{eq:network_output}
\end{equation}
where $\mathcal{F}_{\theta}(\cdot)$ denotes the Transformer-based network with parameters $\theta$, and
\begin{equation}
\mathbf{w}=[w_{0},w_{1},\ldots,w_{N-1}]^{\mathrm T}\in\mathbb{R}^{N}
\label{eq:filter_vector}
\end{equation}
is the generated control filter of length $N=512$.

In implementation, the front-end consists of a Conv1d layer with 1 input channel, 256 output channels, kernel size 64, stride 4, and padding 30, followed by batch normalization, ReLU, and max pooling with stride 4. This results in an overall temporal downsampling factor of 16. Positional encoding with maximum length 912 is added before a Transformer encoder with $d_{\mathrm{model}}=256$, 8 attention heads, 1 encoder layer, feedforward dimension 1024, dropout 0.1, and pre-normalization. The output head consists of Linear$(256\rightarrow512)$, ReLU, Dropout$(0.1)$, and Linear$(512\rightarrow512)$, producing a control filter of length $N=512$. The total number of trainable parameters is 1,201,152.

Compared with previous GFANC methods, the proposed co-processor introduces a Transformer architecture for control-filter generation and directly outputs the final control-filter coefficients. The first design allows the network to model broader temporal dependencies than a purely convolutional co-processor, while the second removes the decomposition--recombination stage and avoids the dependence on an intermediate sub-filter representation. The trade-off is that the network must regress a higher-dimensional target directly, which may potentially require more training data.

\subsection{End-to-end differentiable training}

During training, the co-processor and the ANC forward path are integrated into one differentiable system. Let $\hat{s}(n)$ denote the estimated secondary-path impulse response used in training. The filtered reference is
\begin{equation}
x'(n)=x(n)\ast \hat{s}(n),
\label{eq:filtered_reference}
\end{equation}
where $\ast$ denotes linear convolution. For a generated control filter $\mathbf{w}$, the anti-noise is computed as
\begin{equation}
y(n)=\sum_{k=0}^{N-1} w_k\,x'(n-k),
\label{eq:anti_noise}
\end{equation}
where $w_k$ is the $k$-th control-filter coefficient. The residual error is therefore
\begin{equation}
e(n)=d(n)-\sum_{k=0}^{N-1} w_k\,x'(n-k).
\label{eq:explicit_error}
\end{equation}

For one frame containing $T$ samples, the unsupervised training objective is defined directly from the residual error:
\begin{equation}
\mathcal{L}=\frac{1}{T}\sum_{n=0}^{T-1} e^{2}(n),
\label{eq:loss_basic}
\end{equation}
where $\mathcal{L}$ denotes the training loss. A weighted version can also be used:
\begin{equation}
\mathcal{L}=\frac{1}{T}\sum_{n=0}^{T-1}\alpha_n e^{2}(n),
\label{eq:loss_weighted}
\end{equation}
where $\alpha_n$ is the weighting coefficient at sample $n$. In our implementation, these weights are generated from a forgetting-factor scheme with $\lambda=0.999$.

The key point here is that the mapping
\begin{equation}
\mathbf{x}_f \rightarrow \mathbf{w} \rightarrow y(n) \rightarrow e(n) \rightarrow \mathcal{L}
\label{eq:pipeline}
\end{equation}
is differentiable. Therefore, the network parameters can be updated by backpropagation:
\begin{equation}
\theta \leftarrow \theta-\eta \nabla_{\theta}\mathcal{L},
\label{eq:update}
\end{equation}
where $\eta$ is the learning rate. In contrast to supervised GFANC, no labelled target filters are required; the co-processor is trained directly by minimizing the residual noise after cancellation.

\subsection{Inference and deployment}

After training, only the forward part of the co-processor is retained. In deployment, each buffered reference frame is fed into the Transformer-based network to generate the current control-filter coefficients according to Eq.~(\ref{eq:network_output}). The generated filter is then assigned to the controller and used for sample-wise noise cancellation until the next frame update arrives.

Overall, the proposed framework combines a frame-wise Transformer co-processor with a sampling-rate ANC controller in a unified end-to-end learning architecture. This design enables direct learning of control-filter generation from the residual-noise objective while preserving the practical real-time structure of fixed-filter ANC systems.

\section{Experimental Setup}

This section describes the datasets, acoustic paths, model configurations, baselines, and evaluation protocol used to assess the proposed Transformer-based E2E-CFG method. Because the proposed model is trained only on synthetic noises, its performance on unseen real-world noises is of particular interest.

\subsection{Datasets and acoustic paths}

The proposed model is trained using 83,977 synthetic band-limited noise samples. Each sample has a duration of 1 s and a sampling rate of 13 kHz. The synthetic noises are generated by applying band-pass filters with random center frequencies and bandwidths to white noise, with effective frequency content covering 20--1900 Hz. The dataset is divided into 79,977 training samples, 2,000 validation samples, and 2,000 test samples. During training, additive Gaussian noise with an SNR of 10 dB is further added to the filtered reference signal to simulate sensor noise.

The acoustic paths are also synthetically generated. A band-limited acoustic path covering 10--3000 Hz is used, and the same acoustic path is adopted in both training and testing.

To evaluate generalization ability, the trained model is tested on two groups of unseen noises:
\begin{itemize}[noitemsep]
\item \textbf{Real noises}: aircraft, compressor, genset, handheld drill, large SUV pass-by, mixed aircraft traffic, motorbike, and traffic.
\item \textbf{Synthetic band-limited noises}: 20--490 Hz, 490--960 Hz, 20--960 Hz, and 1430--1900 Hz.
\end{itemize}

\subsection{Model configurations and training hyperparameters}

The proposed Transformer-based model uses the configuration described in Section 2, with input frame length $L=13{,}000$ samples and control-filter length $N=512$. It is trained with Adam, weight decay $10^{-4}$, initial learning rate $5\times10^{-4}$, batch size 128, and 40 epochs. A StepLR scheduler is used with step size 5 and decay factor 0.5.

As a baseline, GFANC uses a Conv1d front-end with 128 output channels, kernel size 80, stride 4, and padding 38, followed by batch normalization, ReLU, max pooling, two residual blocks, adaptive average pooling, and a Linear$(128\rightarrow15)$ layer with sigmoid activation. It uses $M=15$ sub-filters obtained by uniformly partitioning a wideband pre-trained control filter in the frequency domain, with the same control-filter length $N=512$. The initial learning rate is $10^{-2}$, training runs for 10 epochs, and the StepLR scheduler uses step size 3 with decay factor 0.5. The total number of trainable parameters is 211,215.

The proposed method is compared with two baselines:
\begin{itemize}[noitemsep]
\item \textbf{FxNLMS}: the conventional adaptive ANC baseline, with filter length 512 and step size $\mu=0.001$.
\item \textbf{GFANC}: the unsupervised GFANC framework \cite{luo2024unsupervised} with a CNN-based co-processor.
\end{itemize}

The main evaluation metric is the noise reduction (NR) level in dB. For each test noise, the controller is run for 5 s, and the NR is computed over the last 1 s. The reported average NR values are arithmetic means over the corresponding test noises.

\begin{table}[!t]
\caption{Noise reduction (NR) results in dB on unseen real and synthetic noises. All tested real noises and tested synthetic noises are reported. The best result in each row is shown in bold. For each test noise, the controller is run for 5 s and the NR is computed over the last 1 s.}
\label{tab:overall_results}
\centering
\begin{tabular}{l l c c c}
\hline
\textbf{Category} & \textbf{Noise} & \textbf{GFANC} & \textbf{E2E-CFG} & \textbf{FxNLMS} \\
\hline
Real & Aircraft & 15.88 & \textbf{17.83} & 9.17 \\
Real & Compressor & \textbf{21.96} & 19.88 & 14.78 \\
Real & Genset & 12.32 & \textbf{17.03} & 9.01 \\
Real & Handheld drill & 20.65 & \textbf{22.83} & 16.96 \\
Real & Large SUV pass-by & 14.84 & \textbf{17.77} & 9.70 \\
Real & Mix aircraft traffic & 13.03 & \textbf{16.67} & 8.40 \\
Real & Motorbike & \textbf{21.28} & 17.94 & 10.02 \\
Real & Traffic & 13.09 & \textbf{16.90} & 10.96 \\
\hline
\multicolumn{2}{l}{\textbf{Real noise average}} & 16.63 & \textbf{18.36} & 11.13 \\
\hline
Synthetic & 20--490 Hz & \textbf{21.24} & 19.35 & 21.15 \\
Synthetic & 490--960 Hz & 13.07 & 15.32 & \textbf{21.50} \\
Synthetic & 20--960 Hz & 16.23 & \textbf{20.29} & 12.43 \\
Synthetic & 1430--1900 Hz & 14.63 & 19.02 & \textbf{21.14} \\
\hline
\multicolumn{2}{l}{\textbf{Synthetic noise average}} & 16.29 & 18.50 & \textbf{19.06} \\
\hline
\end{tabular}
\end{table}

\section{Results}

\subsection{Comparison on unseen noises}

Table~\ref{tab:overall_results} reports the NR results on all tested real noises and all four tested synthetic noises. On the real-noise set, the proposed Transformer-based E2E-CFG outperforms GFANC in six out of eight cases and achieves the highest average NR of 18.36 dB, compared with 16.63 dB for GFANC and 11.13 dB for FxNLMS. On the synthetic-noise set, the results are more mixed across the three methods, and FxNLMS achieves the highest average NR of 19.06 dB, slightly higher than 18.50 dB for the proposed method and 16.29 dB for GFANC.

These results suggest that the proposed E2E-CFG is more advantageous on unseen real-noise conditions, whereas FxNLMS remains competitive on several synthetic band-limited noises with relatively regular spectral structure.

\begin{figure}[!ht]
  \centering
  \begin{minipage}[t]{0.45\linewidth}
    \centering
    \includegraphics[width=\linewidth]{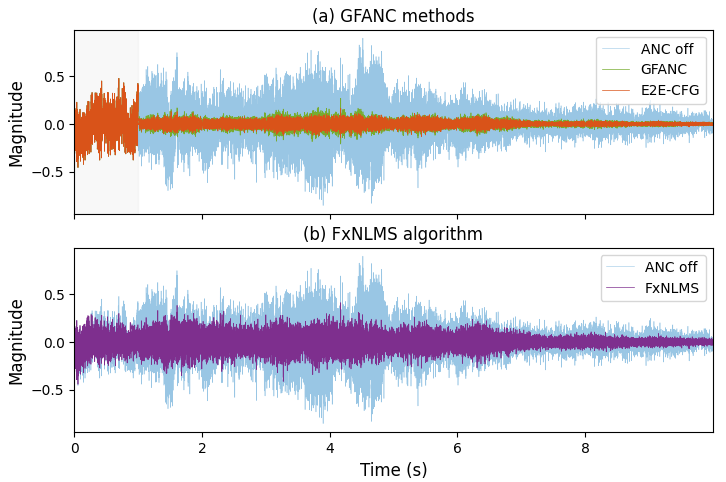}
  \end{minipage}
  \hfill
  \begin{minipage}[t]{0.45\linewidth}
    \centering
    \includegraphics[width=\linewidth]{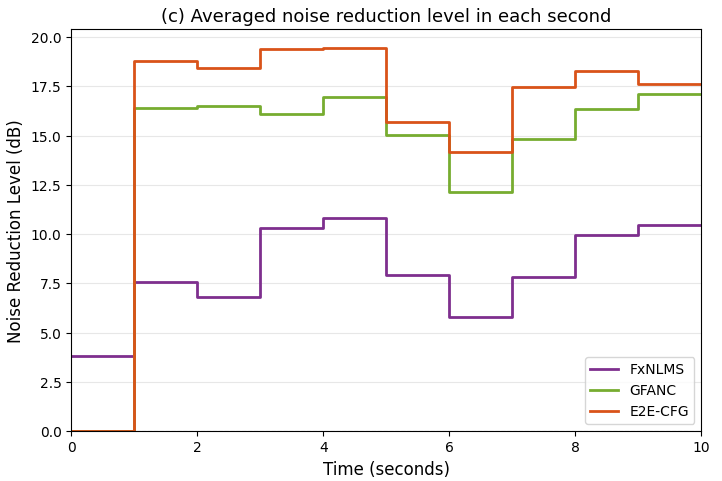}
  \end{minipage}
  \caption{Comparison of the proposed E2E-CFG with GFANC and FxNLMS. The two panels show the waveform-level residual signals and the averaged noise reduction level in each second, respectively.}
  \label{fig:comparison_main}
\end{figure}

\subsection{Time-domain performance}

Figure~\ref{fig:comparison_main} compares the proposed E2E-CFG with GFANC and FxNLMS in both waveform and averaged time-domain views. The waveform comparison shows that both learning-based methods suppress the disturbance more effectively than FxNLMS under the tested case, while the proposed E2E-CFG generally yields lower residual magnitude than GFANC. The averaged noise reduction curve further shows that the proposed method achieves consistently higher noise reduction over most time intervals.

\begin{figure}[!ht]
  \centering
  \includegraphics[width=\linewidth]{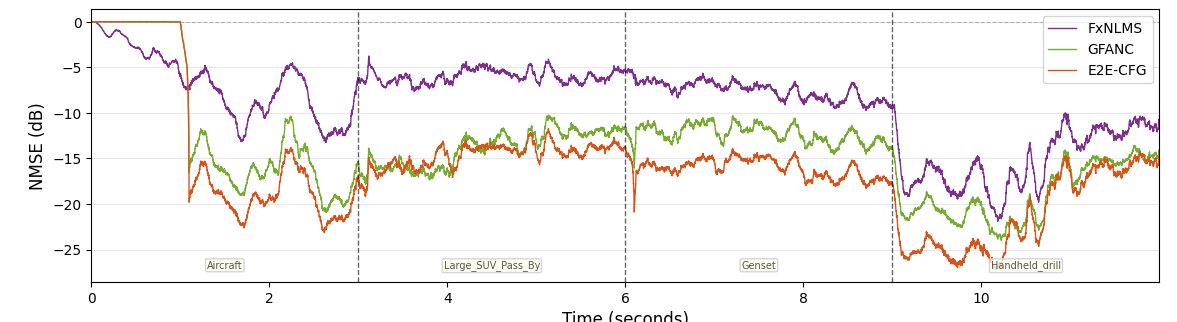}
  \caption{Time-varying NMSE curves under sequential noise-type changes. The test signal contains several real-noise segments with abrupt transitions, including aircraft, large SUV pass-by, genset, and handheld drill.}
  \label{fig:noise_change}
\end{figure}

Figure~\ref{fig:noise_change} further illustrates the behavior of different methods when the noise type changes over time. The proposed E2E-CFG maintains lower NMSE than both GFANC and FxNLMS across most segments, and its advantage remains visible after abrupt switches between different real-noise conditions.

\section{Discussion}

\subsection{Real-noise robustness}

An important observation from Table~\ref{tab:overall_results} is that the proposed method shows its clearest advantage on the real-noise set. In terms of average NR, E2E-CFG achieves 18.36 dB, which is higher than both GFANC and FxNLMS. By contrast, on the synthetic band-limited noises, the three methods are more competitive and FxNLMS achieves the highest average NR. This suggests that the main advantage of the proposed method lies not in uniformly outperforming all baselines on every test case, but in handling unseen real-world noises with stronger nonstationarity and more complex temporal variation.

This observation is also consistent with Fig.~\ref{fig:noise_change}. When the noise type changes over time, the proposed E2E-CFG maintains lower NMSE than both GFANC and FxNLMS across most segments, and its advantage remains visible after abrupt switches between different real-noise conditions. A possible reason is that the proposed method combines Transformer-based sequence modeling with direct control-filter generation. The Transformer-based co-processor can exploit longer-range temporal dependencies within each buffered frame, while the direct prediction of the full control-filter coefficients avoids the intermediate restriction introduced by sub-filter decomposition and recombination. Together, these two design choices appear to be more beneficial for real-world nonstationary noises than for relatively regular synthetic band-limited test cases.

\subsection{Model complexity}

The stronger real-noise performance of the proposed E2E-CFG is accompanied by increased model complexity. Compared with GFANC, the proposed model has a substantially larger parameter count and storage footprint, increasing from about 0.21M to 1.20M trainable parameters and from 876.7 KB to 5.48 MB in model storage. By contrast, the increase in computational cost per frame is more moderate, with the floating-point operations (FLOPs) rising from 385.9 M to 782.5 M. A further breakdown shows that most of the additional computation is introduced by the attention module.

Therefore, the proposed method should be viewed as a performance--complexity trade-off. In the present experiments, the added complexity is associated with stronger results on the real-noise set and more stable behavior under noise-type changes, as shown in Fig.~\ref{fig:noise_change}. Future work may reduce this complexity through lightweight model design or more efficient sequence modeling architectures \cite{gu2023mamba}, while preserving the advantage of direct control-filter generation under complex real-noise conditions.

\section{Conclusion}

This paper presented a Transformer-based End-to-End Control-Filter Generation (E2E-CFG) framework for active noise control. Compared with previous GFANC approaches, the proposed method employs a Transformer-based co-processor for control-filter generation and directly predicts the final control-filter coefficients, without relying on sub-filter decomposition and recombination. This design enables end-to-end unsupervised training that is directly aligned with the physical objective of residual-noise reduction.

Experimental results on unseen real and synthetic noises showed that the proposed method achieved the strongest average performance on the real-noise set among the tested methods. Together with its more stable behavior under noise-type changes, these results suggest that combining Transformer-based sequence modeling with direct control-filter generation is a promising direction for active noise control under more realistic and time-varying noise conditions.

One limitation of the present work is that the proposed model is developed and evaluated under a fixed acoustic path setting. When the system is transferred to a different acoustic environment, the network generally needs to be retrained. Extending the framework to more diverse acoustic environments therefore remains an important direction for future study. Future work will also explore more efficient model designs.



\bibliographystyle{unsrt}
\bibliography{sample}

\end{document}